\documentclass[11pt,american,twoside,a4paper]{article}
\usepackage[T1]{fontenc}
\usepackage[utf8]{inputenc}
\usepackage{latexsym}
\usepackage{amsmath}
\usepackage{bbm}
\usepackage{amssymb}
\usepackage{babel}
\usepackage{amsfonts}
\usepackage{theorem}
\usepackage{epsfig}
\usepackage{enumerate}
\usepackage{color,xcolor}
\usepackage{xspace,needspace,xparse}
\usepackage[active]{srcltx}
\usepackage[colorlinks=true]{hyperref}
\usepackage{fancyhdr}
\colorlet{darkblue}{blue!50!black}
\hypersetup{
	colorlinks,%
	citecolor=darkblue,%
	filecolor=red,%
	linkcolor=darkblue,%
	urlcolor=darkblue,%
	pdfnewwindow=true,%
	pdfstartview={FitH}
}

\numberwithin{equation}{section}

\newcounter{smallarabics}

\newcounter{smallroman}

\newcommand{\ben}{\begin{enumerate}[{\rm (1)}]}
\newcommand{\een}{\end{enumerate}}

\newtheorem{theorem}{Theorem}[section]
\newtheorem{proposition}[theorem]{Proposition}
\newtheorem{lemma}[theorem]{Lemma}
\newtheorem{definition}[theorem]{Definition}

\usepackage[normalem]{ulem}

\AtBeginEnvironment{theorem}{\Needspace{5\baselineskip}}
\AtBeginEnvironment{proposition}{\Needspace{3\baselineskip}}
\AtBeginEnvironment{definition}{\Needspace{5\baselineskip}}
\AtBeginEnvironment{corollary}{\Needspace{5\baselineskip}}
\AtBeginEnvironment{lemma}{\Needspace{5\baselineskip}}
\AtBeginEnvironment{quote}{\Needspace{5\baselineskip}}

\def\cA{{\mathcal A}}  
 \def\cE{{\mathcal E}} 
 \def\cH{{\mathcal H}} 
 \def\cK{{\mathcal K}} \def\cL{{\mathcal L}}
 \def\cN{{\mathcal N}} \def\cO{{\mathcal O}}
\def\cP{{\mathcal P}}  \def\cR{{\mathcal R}}
\def\cS{{\mathcal S}}

\def\rr{{\mathbb R}}

\def\cc{{\mathbb C}}
\def\nn{{\mathbb N}}

\def\tr{{\rm tr}}

\def\e{{\rm e}}
\def\i{{\rm i}}

\def\d{{\rm d}}

\def\bep{\begin{proposition}}
\def\eep{\end{proposition}}
\def\bet{\begin{theoreme}}
\def\eet{\end{theoreme}}
\def\bel{\begin{lemma}}
\def\eel{\end{lemma}}

\newcommand{\ds}{\displaystyle}

\def\Re{{\rm Re}}
\def\Im{{\rm Im}}

\def\one{{\mathchoice {\rm 1\mskip-4mu l} {\rm 1\mskip-4mu l} {\rm 1\mskip-4.5mu l} {\rm 1\mskip-5mu l}}}

\def\qed{\hfill $\Box$\medskip}

\def\textsl{{}}

\def\c0inf{C_0^\infty}

\def\i{{\rm i}}
\newcommand{\beq}{\begin{equation}}
\newcommand{\eeq}{\end{equation}}
\newcommand{\bear}[1]{\begin{array}{#1}}
\newcommand{\ear}{\end{array}}

\def\sgn{{\rm sgn}}

\renewcommand{\i}{\mathrm{i}}

\renewcommand{\d}{\mathrm{d}}

\def\qed{$\Box$\medskip}

\def\bel{\begin{lemma}}
\def\eel{\end{lemma}}
\def\bet{\begin{theoreme}}
\def\eet{\end{theoreme}}
\def\bed{\begin{definition}}
\def\eed{\end{definition}}
\def\bep{\begin{proposition}}
\def\eep{\end{proposition}}

\def\bar{\overline}

\def\Ker{{\rm Ker}\,}
\def\Dom{{\rm Dom}\,}

\def\Ent{{\rm Ent}}

\def\fM{{\mathfrak M}}
\def\fF{{\mathfrak F}}

\def\dim{{\rm dim}}

\def\fr{{\rm fr}}

\begin{document}
\def\today{}
\title{A note on  two-times measurement entropy production and modular theory}
\author{T. Benoist$^{1}$, L. Bruneau$^{2}$, V. Jak\v{s}i\'c$^{3,4}$, A. Panati$^5$, C.-A. Pillet$^5$
\\ \\
$^1$ Institut de Math\'ematiques de Toulouse, UMR 5219\\
Universit\'e de Toulouse, CNRS, UPS\\
31062 Toulouse Cedex 9, France
\\ \\
$^2$ Department of Mathematics\\
CY Cergy Paris University, CNRS UMR 8088\\
2 avenue Adolphe Chauvin, 95302 Cergy-Pontoise, France
\\ \\
$^3$ Centre de Recherches Math\'ematiques -- CNRS UMI 3457 \\
Universit\'e de Montr\'eal\\
Montr\'eal, QC, H3C 3J7, Canada
\\ \\
$^4$Department of Mathematics and Statistics, McGill University\\
805 Sherbrooke Street West, Montreal,  QC,  H3A 2K6, Canada
\\ \\
$^5$Universit\'e de Toulon, CNRS, CPT, UMR 7332, 83957 La Garde, France\\
Aix-Marseille Univ, CNRS, CPT, UMR 7332, Case 907, 13288 Marseille, France
}
\maketitle
\thispagestyle{empty}

\noindent{\small{\bf Abstract.}
Recent theoretical investigations of  the two-times measurement  entropy
production (2TMEP) in quantum statistical mechanics have shed a new light on the
mathematics and physics  of the quantum-mechanical probabilistic rules. Among
notable developments are the extensions of  entropic fluctuation relations  to
the quantum domain and the discovery of a  deep link between 2TMEP and modular
theory of operator algebras. All these developments concerned the setting where
the state of the system at the  instant of the first measurement is the same as
the state whose entropy production is measured. In this work we consider the
case where these two states are different and link this more general 2TMEP to
modular theory. The established connection allows us to show  that under
general ergodicity assumptions the  2TMEP is essentially independent of  the
choice of the system state at the instant of the first measurement due to a
decoherence  effect induced by the first measurement.  This stability  sheds a
new  light on the concept of quantum entropy production, and, in particular, on
possible quantum formulations of the celebrated classical Gallavotti--Cohen
Fluctuation Theorem which will be studied in a continuation of this work.
}

\tableofcontents

\section{Introduction}
\label{sec-intr-n}

Starting with the seminal work~\cite{Haag1967}, the mathematical theory of
equilibrium quantum statistical mechanics based on the KMS-condition has
developed rapidly in 1970's, resulting in a structure of rare unity and beauty
summarized in the  classical monographs~\cite{Bratteli1987,Bratteli1981}. A
large part of these developments was centered around the link between the
KMS-condition and the modular theory of operator algebras.

Modular theory has also played a central role in the more recent developments in
non-equilibrium quantum statistical mechanics initiated
in~\cite{Jaksic2001a,Ruelle2001}, where the definition of an entropy production
observable and the related entropy balance equation are given in terms of basic
objects of modular theory;\footnote{See~\cite{Pusz1978} for a pioneering work on
the subject.} a non-exhaustive list of  related works
is~\cite{Salem2007,AbouSalem2007,Ho2000,Aschbacher2003,Aschbacher2006,Aschbacher2007,Froehlich2003a,Froehlich2003,Jaksic2006f,Jaksic2006b,Jaksic2013,Jaksic2006c,Jaksic2006a,Jaksic2007a,Jaksic2010b,Jaksic2012,Jaksic2002b,Jaksic2002a,Jaksic2007b,Merkli2007a,Merkli2007,Matsui2003,Ogata2004,Ojima1989,Ojima1991,Ojima1988,Pillet2001,Ruelle2002,Tasaki2006,Tasaki2003,Tasaki2005}.

Perhaps more surprising were parallel developments related to the search for a
quantum extension of the celebrated fluctuation relations of classical
non-equilibrium statistical
mechanics~\cite{Evans1994,Gallavotti1995b,Gallavotti1995c}, see also the
review~\cite{Jaksic2011}. The  first of them introduced the two-times
measurement entropy production~\cite{Kurchan2000,Tasaki2000}. The spectral
measure of a relative modular operator was the central object of the second
one~\cite{Tasaki2003}.\footnote{Another early work on the subject
is~\cite{DeRoeck2009},}  These two proposals turned out to be equivalent,
shedding an unexpected light on both quantum mechanical probabilistic rules and
modular theory. A pedagogical discussion  of this topic can be found in the
lecture notes~\cite{Jaksic2010b}.

In this work we continue to study the link between the two-times measurement
entropy production and modular theory. The more general setting we consider
concerns the choice of the system state  at the instant of the first
measurement, which is here assumed to be arbitrary. Somewhat surprisingly, under
mild ergodicity assumptions, the modular link we establish gives that the
two-times measurement entropy production in open quantum systems is essentially
independent of the  state of the system at the instant of the first measurement.
This  stability result will be the  starting point of our follow-up
work~\cite{Benoist2024}, in which we propose an extension of the celebrated
Gallavotti--Cohen Fluctuation Theorem~\cite{Gallavotti1995b,Gallavotti1995c} to
the quantum domain; see also Remark 4 in Section~\ref{sec-remarks}.

This note is organized as follows. For notational purposes, the elements of
algebraic quantum statistical mechanics and modular theory that we will need are
briefly reviewed in Sections~\ref{sec-alg} and~\ref{sec-modular-n}. The
two-times measurement entropy production of finite quantum system is discussed
in Section~\ref{sec-modular-n}. Our  results are stated in
Section~\ref{sec-mainresults} and are briefly discussed in
Section~\ref{sec-remarks}. The proofs are given in Section~\ref{sec-proofs}.

\paragraph*{Acknowledgments} The work of CAP and VJ  was partly  funded by  the
CY Initiative grant "Investissements d'Avenir", grant number ANR-16-IDEX-0008.
The work of TB was funded by the ANR project  “ESQuisses”, grant number
ANR-20-CE47-0014-01, and  by the ANR project  “Quantum Trajectories”, grant
number ANR-20-CE40-0024-01. VJ acknowledges the support of NSERC. A part of this
work was done during long term visits of AP and LB to McGill University and the
CRM-CNRS International Research Laboratory IRL3457 at the Université de Montréal.
The visits of AP were supported by the CRM Simons and FRQNT-CRM-CNRS programs, and that of
LB by CNRS.

\subsection{Algebraic quantum statistical mechanics}
\label{sec-alg}

We start with the setting of a quantum system with finite dimensional Hilbert
space $\cK$. We will refer to such quantum systems as {\em finite}. $\cO_\cK$
denotes the $C^\ast$-algebra of all linear maps $A:\cK\to\cK$ and
$\cS_\cK\subset\cO_\cK$ the set of all density matrices on $\cK$.
Observables of the system are identified with elements of $\cO_\cK$ and
states with elements of $\cS_\cK$, with the usual duality $\nu(A)=\tr(\nu A)$,
$\nu\in\cS_\cK$, $A\in \cO_\cK$. The number $\nu(A)$ is interpreted as
the expectation value of the observable $A$ when the system is in the state
$\nu$. A state $\nu$ is called faithful if $\nu>0$. The dynamics is described
by the system Hamiltonian $H=H^\ast\in\cO_\cK$ and the induced group
$\tau=\{\tau^t\mid t\in\rr\}$ of $\ast$-automorphisms of $\cO_\cK$
defined by
\[
\tau^t(A)=\e^{\i tH}A\e^{-\i tH}.
\]
We will sometimes write $A_t$ for $\tau^t(A)$ and call the map $A\mapsto A_t$
the Heisenberg picture dynamics. In the dual Schrödinger picture the states
evolve in time as $\nu\mapsto\nu_t$ where
\[
\nu_t=\e^{-\i tH}\nu\e^{\i tH}.
\]
Obviously, $\nu_t(A)=\nu(A_t)$. The time-correlations are quantified by the
function
\beq
F_{\nu, A, B}(t)=\nu(AB_t).
\label{t-cor}
\eeq
A triple $(\cO_\cK,\tau,\omega)$, where $\omega$ is the reference state of the
system, is called a  finite quantum dynamical system. This system  is said to be
in thermal equilibrium at inverse temperature $\beta\in\rr$ if its reference
state is the Gibbs canonical ensemble
\beq
\omega=\frac{\e^{-\beta H}}{\tr (\e^{-\beta H})}.
\label{gibbs-n}
\eeq
The Gibbs ensemble~\eqref{gibbs-n} is the unique state $\nu\in\cS_\cK$ satisfying
the KMS relation
\[
\nu(AB_{t+\i\beta})=\nu(B_tA)
\]
for all $A,B\in\cO_\cK$ and $t\in\rr$.

More generally, in algebraic quantum statistical mechanics observables are
described by elements of a $C^\ast$-algebra $\cO$ with identity $\one$. For a
large part of the general theory, no other structure is imposed on $\cO$. States
are elements of $\cS_\cO$, the set of positive normalized\footnote{$\nu(A^\ast
A)\ge0$ for all $A\in\cO$ and $\nu(\one)=1$.} elements $\nu$ of the dual
$\cO^\ast$ of $\cO$. The number $\nu(A)$ is interpreted as the expectation value
of the observable $A$ when the system is in the state $\nu$.

The Heisenberg picture dynamics is described by a strongly
continuous\footnote{$\lim_{t\to0}\|\tau^t(A)-A\|=0$ for all $A\in\cO$.}
group $\tau=\{\tau^t\mid t\in\rr\}$ of $\ast$-automorphisms of $\cO$.
The group $\tau$ is called $C^\ast$-dynamics and the pair $(\cO,\tau)$
a $C^\ast$-dynamical system. The dual group $\tau^\ast$ preserves $\cS_\cO$
and describes the Schrödinger picture dynamics. We  write $A_t$ for
$\tau^t(A)$, $\nu_t$ for $\tau^{t\ast}(\nu)=\nu\circ\tau^t$, and use the same
shorthand~\eqref{t-cor} for time-correlation functions. A state $\nu$ is called
$\tau$-invariant (or stationary) if $\nu_t=\nu$ for all $t\in\rr$. The set of all
$\tau$-invariant states is denoted by $\cS_\tau$ and is always non-empty. A
triple $(\cO,\tau,\omega)$, where $\omega$ is the reference state of the system,
is called $C^\ast$-quantum dynamical system. A state $\omega\in\cS_\tau$ is called
ergodic if
\[
\lim_{T\to\infty}\frac{1}{2T}\int_{-T}^T\omega(B^\ast A_tB)\d t=\omega(A)\omega(B^\ast B)
\]
holds for all $A,B\in\cO$.

Time-reversal plays an important role in statistical mechanics. An anti-linear
involutive $\ast$-automorphism $\Theta$ of $\cO$ is called time-reversal of
$(\cO,\tau)$ if
\[
\Theta\circ\tau^t=\tau^{-t}\circ\Theta
\]
for all $t\in\rr$. A state $\omega$ is called time-reversal invariant if there
exists a time-reversal $\Theta$ such that $\omega\circ\Theta(A)=\omega(A^\ast)$
for all $A\in\cO$.

For $\beta\in\rr^\ast$, $\nu\in\cS_\cO$ is called $(\tau,\beta)$-KMS state if, for all
$A,B\in\cO$, the function $\rr\ni t\mapsto F_{\nu,A,B}(t)$ has an analytic
extension to the strip $0<\sgn(\beta)\Im z<|\beta|$ that is bounded and
continuous on its closure and satisfies the KMS-boundary condition
\[
F_{\nu,A,B}(t+\i\beta)=\nu(B_tA)
\]
for all $t\in\rr$. We denote by $\cS_{(\tau,\beta)}$ the set of all
$(\tau,\beta)$-KMS states. At the current level of generality this set might be
empty. One always has $\cS_{(\tau,\beta)}\subset\cS_\tau$. A $C^\ast$-quantum
dynamical system $(\cO,\tau,\omega)$ is said to be in thermal
equilibrium at inverse temperature $\beta\in\rr^\ast$ (or just thermal) if
$\omega\in\cS_{(\tau,\beta)}$.

A state $\nu$ is called modular if there exists a $C^\ast$-dynamics
$\varsigma_\nu$ on $\cO$ such that $\nu\in\cS_{(\varsigma_\nu,-1)}$.
$\varsigma_\nu$ is called modular group of $\nu$ and is unique when it exists.
We denote by $\delta_\nu$ the generator of $\varsigma_\nu$ with the convention
$\varsigma_\nu^t=\e^{t \delta_\nu}$. If $\nu\in\cS_{(\tau,\beta)}$, then it is
modular and its modular group is $\varsigma_\nu^t=\tau^{-\beta t}$
(or equivalently, $\delta_\nu=-\beta\delta$, where $\delta$ is the generator of
$\tau$).

A special class of quantum dynamical systems, the so-called open quantum
systems, play a privileged role in the study of non-equilibrium quantum
statistical mechanics, and we proceed to describe them.

Consider $M$ thermal reservoirs $\cR_j$ described by $C^\ast$-quantum dynamical
systems $(\cO_j,\tau_j,\omega_j)$. We denote by $\delta_j$ the generator
of $\tau_j$. The reservoir $\cR_j$ is assumed to be in thermal equilibrium at
inverse temperature $\beta_j>0$, that is, we assume that $\omega_j$ is a
$(\tau_j,\beta_j)$-KMS state on $\cO_j$. In the absence of interaction,
the combined  reservoir system $\cR=\cR_1+\cdots+\cR_M$ is
described by the quantum dynamical system  $(\cO_\cR,\tau_\cR,\omega_\cR)$,
where\footnote{Whenever the meaning is clear within the context,
we write $A$ for $A\otimes\one$ and $\one\otimes A$, $\delta_j$ for
$\delta_j\otimes\mathrm{Id}$, $\mathrm{Id}\otimes\delta_j$, etc.}
\begin{align*}
\cO_\cR&=\cO_1\otimes\cdots\otimes\cO_M,\\[1mm]
\tau_\cR&=\tau_1\otimes\cdots\otimes\tau_M,\\[1mm]
\omega_\cR&=\omega_1\otimes\cdots\otimes\omega_M.
\end{align*}
We will consider two kinds of systems: directly coupled reservoirs and
reservoirs coupled through a small system ${\cal S}$, with Hilbert space
$\cK_\cS$. With a slight abuse of terminology, we will refer to both of them as
{\sl open quantum systems}.

In the first case, the interaction is described by a self-adjoint $V\in\cO_\cR$
and the interacting dynamics $\tau$ is generated by $\delta=\delta_\cR
+\i[V,\,\cdot\;]$, where $\delta_\cR=\sum_j\delta_j$ is the generator of
$\tau_\cR$.

In the second case, let $(\cO_\cS,\tau_\cS,\omega_\cS)$ be the finite
dimensional $C^\ast$-quantum dynamical system describing $\cS$\footnote{We
abbreviated by $\cO_\cS$ the $C^\ast$-algebra $\cO_{\cK_\cS}$ of all
linear operators on $\cK_\cS$.}, where we assume that $\omega_\cS>0$. The
generator of $\tau_\cS$ is $\delta_\cS=\i[H_\cS,\,\cdot\;]$, where $H_\cS$ is the
Hamiltonian of $\cS$. In the absence of interaction, the joint system $\cS+\cR$
is described by the $C^\ast$-quantum dynamical system
$(\cO,\tau_\fr,\omega)$ where
\[
\cO=\cO_\cS\otimes\cO_\cR,\qquad
\tau_\mathrm{fr}=\tau_\cS\otimes\tau_\cR,\qquad
\omega=\omega_\cS\otimes\omega_\cR.
\]
The state $\omega$ is obviously modular. The interaction of $\cS$ with $\cR_j$
is described by a self-adjoint element $V_j\in \cO_\cS\otimes\cO_j$, and  the full
interaction by $V=\sum_jV_j$. The interacting dynamics $\tau$ is generated
by $\delta=\delta_\cS+\delta_\cR+\i [V,\,\cdot\;]$. In what follows, we will always take
\beq
\omega_\cS=\frac{\one}{\dim\,\cK_\cS}
\label{s-choice}
\eeq
for the reference state of $\cS$. This choice is made for convenience. It is
easy to show that none of our results depend on a specific choice of $\omega_\cS$
as long as $\omega_\cS>0$; see Remark~6 in Section~\ref{sec-remarks}.

The above description of open quantum system is sometimes  modified in the case
of fermionic systems. The modifications are straightforward, and they do not
affect any of our results; see~\cite{Aschbacher2006,Jaksic2007a}.

Modular theory and the closely related Araki's perturbation theory of
KMS-structure play a central role in algebraic  quantum statistical mechanics.
A basic introduction to this subject can be found
in~\cite{Bratteli1987,Bratteli1981}; see also~\cite{Derezinski2003a} and
references therein for modern expositions. A pedagogical introduction to
modular theory in the context of finite quantum systems can be found
in~\cite{Jaksic2010b}. We will not  give a detailed review of modular theory
in this paper and only a short introduction to basic notions will be presented
in Section~\ref{sec-modular-n}. However, as we proceed with the proofs, we will
give references to the results we will use.

$W^\ast$-dynamical systems play a distinguished role in modular theory. Consider
a pair $(\fM,\tau)$ where $\fM$ is $W^\ast$-algebra and
$\tau=\{\tau^t\mid t\in\rr\}$ is a pointwise $\sigma$-weakly continuous group of
$\ast$-automorphisms on $\fM$. We shall refer to such $\tau$ as
$W^\ast$-dynamics. A triple $(\fM,\tau,\omega)$, where $\omega$ is a normal
state on $\fM$, is called a $W^\ast$-dynamical system. In the general
development of non-equilibrium quantum statistical mechanics, the
$C^\ast$-quantum dynamical systems are preferred starting point since the
central notion of non-equilibrium steady states cannot be naturally defined in
the $W^\ast$-setting.

\subsection{GNS-representation and modular structure}
\label{sec-mod}
\label{sec-modular-n}

Let $\omega$ be a modular state on $\cO$. We denote by $(\cH_\omega,\pi_\omega,\Omega_\omega)$
the GNS-representation of $\cO$ associated to $\omega$, and by
$\fM_\omega=\pi_\omega(\cO)^{\prime\prime}$ the enveloping  von~Neumann
algebra of bounded  operators on $\cH_\omega$. In what follows, we drop
the subscript $\omega$ whenever the meaning is clear within the context. Since
the state $\omega$ is  assumed to be modular, the cyclic vector $\Omega$ is
separating for $\fM$\footnote{See~\cite[Corollary 5.3.9]{Bratteli1981}}, and in
particular $\|\pi(A)\|=\|A\|$ for all $A\in\cO$. Whenever the meaning is clear
within the context, we will denote $\pi(A)$ by $A$.

$\cN$ denotes the set of all normal states on $\fM$, {\sl i.e.,} the
states described by density matrices on $\cH$. Obviously, an element
of $\cN$ also defines a state on $\cO$ and any state  on $\cO$
that arises in this way is called $\omega$-normal. Again, whenever the meaning
is clear within the context, we will denote such states by the same letter. In
particular, the vector state $\fM\ni A\mapsto\langle\Omega,A\Omega\rangle$
is  denoted by $\omega$.

We will assume that the reader is familiar with the basic notions of
Tomita-Takesaki's modular theory; see any of the
references~\cite{Bratteli1987,Bratteli1981,Derezinski2003a,Haag1996,Ohya1993,Pillet2006,Stratila1981}.
For definiteness, we will use the same notation and terminology as
in~\cite[Section 5]{Jaksic2012}. $\cH^+$ and $J$ denote the natural cone and
modular conjugation associated to the pair $(\fM,\Omega)$. The unique vector
representative of $\nu\in\cN$ in the natural cone is denoted by
$\Omega_\nu$. The modular operator of $\nu\in\cN$ is denoted by
$\Delta_\nu$. The relative modular operator of a pair $(\nu,\rho)$ of
$\omega$-normal states is denoted by $\Delta_{\nu|\rho}$.

The relative entropy of a pair $(\nu,\rho)$ of $\omega$-normal faithful states is
 \[
{\rm Ent}(\nu|\rho)=\langle\Omega_\nu,\log\Delta_{\rho|\nu}\Omega_\nu\rangle.
 \]
This is the original definition of Araki~\cite{Araki1975/76,Araki1977}, with
the sign and ordering convention of~\cite{Jaksic2001a}. In particular,
${\rm Ent}(\nu|\rho)\leq0$ with equality iff $\rho=\nu$. For additional
information about relative entropy we refer the reader to~\cite{Ohya1993}.

Since $\omega$ is $\varsigma_\omega$ invariant, the family
$\{\pi\circ\varsigma_\omega^t\mid t\in\rr\}$ extends to a $W^\ast$-dynamics on $\fM$
which we again denote by $\varsigma_\omega$. For $A\in\fM$, one has
\[
\varsigma_\omega^t(A)=\Delta_\omega^{\i t}A\Delta_\omega^{-\i t}.
\]
More generally, to any faithful $\nu\in\cN$ one associates the
$W^\ast$-dynamics $\varsigma_\nu$ given by
\[
\varsigma_\nu^t(A)=\Delta_\nu^{\i t}A\Delta_\nu^{-\i t}.
\]
$\varsigma_\nu$ is called the modular dynamics of $\nu$,
and $\nu$ is a $(\varsigma_\nu,-1)$-KMS state on $\fM$.

Throughout the paper we assume that the following holds:

\begin{quote}{\bf (Reg1)} The family $\{\pi\circ\tau^t\mid t\in\rr\}$  extends the
interacting dynamics $\tau$ to a $W^\ast$-dynamics on $\fM$ which we again denote
by $\tau$.
\label{reg1}
\end{quote}
\newcommand{\RegOne}{{\hyperref[reg1]{{\rm (Reg1)}}}}%

This assumption is automatically satisfied by the interacting dynamics of the
two kinds of open quantum systems introduced in the previous section. It ensures
that $\omega_t$ is a $\omega$-normal faithful state on $\fM$ and that there
exists a unique self-adjoint operator $\cL$ on $\cH$, called the standard
Liouvillean of $\tau$, such that
\[
\tau^t(A)=\e^{\i t\cL}A\e^{-\i t\cL},\qquad
\e^{-\i t\cL}\cH^+=\cH^+,
\]
for all $A\in\fM$ and $t\in\rr$. The vector representative of $\omega_t$ in
$\cH^+$ is  $\e^{-\i t\cL}\Omega$. We will make use of the following
well-known result.
\begin{theorem}
\ben
\item $\omega \in {\cal S}_\tau \Leftrightarrow {\cal L}\Omega=0$.
\item Suppose that  $\omega \in {\cal S}_\tau$. Then the quantum dynamical system $(\cO, \tau, \omega)$ is ergodic iff\,  $0$ is a simple eigenvalue of ${\cal L}$.
\een
\label{thm-li}
\end{theorem}

For latter reference we also recall the following well-known result that
identifies ergodicity with the so-called property of return to
equilibrium~\cite{Robinson1973}.

\begin{theorem}  Suppose that $\omega\in\cS_\tau$. Then the quantum dynamical
system $(\cO,\tau,\omega)$ is ergodic iff for any $\omega$-normal state
$\nu$ and all $A\in\cO$ one has
\[
\lim_{T\to\infty}\frac{1}{T}\int_0^T\nu(\tau^t(A))\d t=\omega(A).
\]
\label{thm-ret-erg}
\end{theorem}

To any pair of faithful $\omega$-normal states $\nu$ and $\rho$ one associates
the Connes cocycle
\beq
[D\nu:D\rho]_{\i t}=\Delta_{\nu|\rho}^{\i t}\Delta_\rho^{-\i t}, \qquad (t\in \rr),
\label{equ:ConDef}
\eeq
which is obviously a family of unitary operators. Its basic property is that
$[D\nu:D\rho]_{\i t}\in\fM$; for additional properties of Connes' cocycle,
see~\cite[Appendix C]{Araki1982}.

The Connes cocycle
\[
[D\omega_t: D\omega]_\alpha, \qquad \alpha \in \i\rr,
\]
will play a central role in our work. In what follows, we assume
\begin{quote}{\bf (Reg2)} For all $t\in\rr$ and $\alpha\in\i\rr$,
\[
[D\omega_t:D\omega]_{\alpha}\in\pi(\cO).
\]
\label{reg2}
\end{quote}
\newcommand{\RegTwo}{{\hyperref[reg2]{{\rm (Reg2)}}}}%
$\pi^{-1}([D\omega_t:D\omega]_{\alpha})$ will be also denoted by
$[D\omega_t:D\omega]_{\alpha}$. For the two kinds of open quantum systems
introduced in the previous section, \RegTwo{} holds if $V\in\Dom(\delta_\omega)$.
This follows from the definition~\eqref{equ:ConDef}, the fact that
$$
\Delta_{\omega_t|\omega}^\alpha=\e^{\alpha(\log\Delta_\omega+\pi_\omega(Q_t))},\qquad
Q_t=\int_0^t\tau^{-s}(\delta_\omega(V))\d s,
$$
as established in the proof of~\cite[Theorem~1.1]{Jaksic2003}, and
time-dependent perturbation theory.

\subsection{Two-times measurement entropy production in finite quantum systems}
\label{sec-two-times}

This notion of entropy production goes back to~\cite{Kurchan2000,Tasaki2000} and
has been studied in detail in~\cite{Jaksic2010b}. Consider a finite quantum
dynamical system on a Hilbert space $\cK$ with Hamiltonian $H$. The measurement
protocol is defined with respect to two faithful states $\nu$ and $\omega$. The
first one, $\nu$, is the state of system at the instant of the first
measurement, and will be a variable in our work. The second one, $\omega$, is
assumed to be fixed and defines the observable to be measured. More precisely,
we will consider two consecutive measurements of the observable
$$
S=-\log\omega
$$
interpreting the increase $\Delta S$ in the outcomes of these two measurement as
the entropy produced by the system during the time interval between the two
measurements. To motivate this interpretation, let the small system $\cS$, with
$\omega_\cS$ given by~\eqref{s-choice}, be coupled to thermal reservoirs
$\cR_1,\ldots,\cR_M$ at inverse temperatures $\beta_1,\ldots,\beta_M$. Setting
$$
\omega=\omega_\cS\otimes\left(\bigotimes_{j=1}^M\frac{\e^{-\beta_jH_j}}{\tr\,\e^{-\beta_jH_j}}\right)
$$
where $H_j$ denotes the Hamiltonian of the $j^\mathrm{th}$ reservoir, we get
$$
\Delta S=\sum_{j=1}^M\beta_j\Delta E_j
$$
where $\Delta E_j$ is the change in the energy of the $j^\mathrm{th}$ reservoir. Thus, $\Delta S$ can be
identified with the entropy dumped by the system $\cS$ in the reservoirs.

Let $\cA$ be a finite alphabet indexing the distinct eigenvalues
$(\lambda_a)_{a\in\cA}$ of $\omega$, and let $P_a$  be the eigenprojection
corresponding to $\lambda_a$. The observable to be measured is described by
the partition of unity $(P_a)_{a\in\cA}$ on $\cK$, with  outcomes of the
measurement labeled by the letters of $\cA$. The two-times measurement protocol
goes as follows. At the instant of the first measurement, when the system is in
the state $\nu$, the outcome $a\in\cA$ is observed with probability
\[
p_\nu(a)=\nu(P_a).
\]
After the measurement the system is in the reduced state
$$
\frac{1}{p_\nu(a)}P_a\nu P_a,
$$
which evolves under the system dynamics over the time interval of length $t$ to
\[
\frac{1}{p_\nu(a)}\e^{-\i tH}P_a\nu P_a\e^{\i tH}.
\]
The second measurement, performed at the end of this time interval,
yields the outcome $b\in\cA$ with probability
$$
p_{\nu,t}(b|a)=\frac{1}{p_\nu(a)}\tr(\e^{-\i tH}P_a\nu P_a\e^{\i tH}P_b).
$$
Finally, the probability of observing the pair $(b,a)$ in this  two-times measurement protocol is
\beq
p_{\nu,t}(b,a)=p_{\nu,t}(b|a)p_\nu(a)=\tr(\e^{-\i tH}P_a\nu P_a\e^{\i tH} P_b).
\label{ss-n}
\eeq
The formula~\eqref{ss-n} defines a probability measure on $\cA\times\cA$ even
when $\nu$ is not faithful, and from now on we drop this
restriction.\footnote{For non-faithful $\nu$ the protocol can be implemented in
a limiting sense by considering a sequence of faithful $\nu_n$'s such that
$\lim_n\nu_n=\nu$.}

The entropy production random variable $\cE:\cA\times\cA\to\rr$ is defined by
$$
\cE(b,a)=-\log\lambda_b+\log \lambda_a,
$$
and its probability distribution with respect to $p_{\nu,t}$ is denoted by $Q_{\nu,t}$,
\[
Q_{\nu,t}(s)=\sum_{\cE(b,a)=s}p_{\nu,t}(b,a).
\]
The statistics of two-times measurement entropy production is described by the
family $(Q_{\nu, t})_{t>0}$. In the case $\nu=\omega$, this family of
probability measures was studied in detail in~\cite{Jaksic2010b}. To the best of
our knowledge, the case $\nu\neq\omega$ was not considered before in the
mathematical physics literature.

We set
\[
\fF_{\nu,t}(\alpha)=\int_\rr\e^{-\alpha s}\d Q_{\nu,t}(s), \qquad (\alpha\in\cc).
\]
The definition of $Q_{\nu,t}$ gives
\[
\fF_{\nu,t}(\alpha)=\tr(\omega_{-t}^\alpha\omega^{-\alpha}\bar\nu)
\]
where $\bar\nu=\sum_{a\in\cA}P_a\nu P_a$. Taking $\nu=\omega$ leads to the formulas
$$
\fF_{\omega,t}(\alpha)=\omega([D\omega_{-t}:D\omega]_\alpha)
=\langle\Omega_\omega,\Delta_{\omega_{-t}|\omega}^{\alpha}\Omega_\omega\rangle
$$
and to the identification of $Q_{\omega,t}$ with the spectral measure of the
operator $-\log\Delta_{\omega_{-t}|\omega}$ for the vector $\Omega_\omega$.
This deep link between the statistics of the two-times measurement entropy
production and modular theory has a somewhat unusual history and  was
discussed in detail in~\cite{Jaksic2010b}.

The  starting point of this work is the observation that for general $\nu$ one
can also  link $Q_{\nu,t}$ to the modular structure via the formula
\beq
\fF_{\nu,t}(\alpha)=\lim_{R\to\infty}\frac{1}{R}\int_0^R
\nu\left(\varsigma_\omega^\theta\left(
[D\omega_{-t}:D\omega]_{\alpha}\right)\right)\d\theta
\label{st-cyr}
\eeq
that follows by an elementary computation. This modular representation of
$\fF_{\nu,t}$ for general $\nu$, to the best of our knowledge, has not
appeared previously in the literature.

We are now ready to state our main results.

\subsection{Main results}
\label{sec-mainresults}

Throughout this section $(\cO,\tau,\omega)$ is a fixed $C^\ast$-quantum
dynamical system with modular reference state $\omega$. Recall that Assumptions
\RegOne{} and \RegTwo{} are in force throughout the paper.

\begin{theorem}\label{gen-ttm-quant-1-n}
For  all $\nu\in\cN$, $t\in\rr$, and $\alpha\in\i\rr$, the limit
\beq
\fF_{\nu,t}(\alpha)=\lim_{R\to\infty}\frac{1}{R}\int_0^R
\nu\left(\varsigma_\omega^\theta\left([D\omega_{-t}:D\omega]_{\alpha}\right)\right)\d\theta
\label{emm-n}
\eeq
exists, and there exists unique Borel probability measure $Q_{\nu,t}$ on $\rr$ such that
\beq
\fF_{\nu,t}(\alpha)=\int_\rr\e^{-\alpha s}\d Q_{\nu,t}(s).
\label{han-ajde}
\eeq
\end{theorem}
The family $(Q_{\nu,t})_{t>0}$ describes the statistics of two-times
measurement entropy production of $(\cO,\tau,\omega)$ with respect to
$\nu$ in the above general setting. This definition, that arises  by modular
extension of the finite quantum system physical notion discussed in the
previous section, requires a general comment in relation to  the thermodynamic
limit procedure; see Remark~1 in Section~\ref{sec-remarks}
and~\cite[Chapter~5]{Jaksic2010b}.

In the case $\nu=\omega$,
\[
\omega\left(\varsigma_\omega^\theta\left([D\omega_{-t}:D\omega]_{\alpha}\right)\right)
=\langle\Omega_\omega,\Delta_{\omega_{-t}|\omega}^{\alpha}\Omega_\omega\rangle
\]
for all $\theta$, and so $Q_{\omega,t}$ is the spectral measure of
$-\log\Delta_{\omega_{-t}|\omega}$ for the vector $\Omega_\omega$. Thus,
$Q_{\omega,t}$ coincides with the proposal of~\cite{Tasaki2003}, where the
authors, unaware of the works~\cite{Kurchan2000,Tasaki2000}, were searching for
a quantum version of the fluctuation relation of classical statistical mechanics;
see Proposition~\ref{qes-1-gen} below. That, in the finite quantum system setting,
this spectral measure coincides with the two-times measurement entropy
production statistics discussed in the previous section did not appear in print
until~\cite{Jaksic2010b}. The basic properties of $Q_{\omega,t}$ are summarized
in
\begin{theorem}\label{gen-ttm-quant}
\ben
\item $\int_\rr s\,\d Q_{\omega,t}(s)=-\Ent(\omega_t|\omega)$. In particular,
$\int_\rr s\,\d Q_{\omega,t}(s)\geq0$ with the equality iff $\omega=\omega_t$.
\item The map $\i\rr\ni\alpha\mapsto\fF_{\omega,t}(\alpha)$ has an analytic
extension to the vertical strip $0<\Re\,\alpha<1$ that is bounded and
continuous on its closure and satisfies
$$
\fF_{\omega,t}(\alpha)=\overline{\fF_{\omega,-t}(1-\bar\alpha)}
$$
for $0\leq\Re\,\alpha\leq 1$ and $t\in\rr$.
\een
In the remaining statements we assume that $\omega$ is time-reversal invariant.
\ben
\setcounter{enumi}{2}
\item For any $\alpha$ satisfying $0\leq\Re\,\alpha\leq 1$,
\beq
\fF_{\omega,t}(\alpha)=\overline{\fF_{\omega,t}(1-\bar \alpha)}.
\label{qes-1-gen}
\eeq
\item Let ${\mathfrak r}:\rr\rightarrow\rr$ be the reflection
${\mathfrak r}(s)=-s$ and $\bar Q_{\nu,t}=Q_{\nu,t}\circ{\mathfrak r}$.
Then the measures $Q_{\omega,t}$ and $\bar Q_{\omega,t}$ are equivalent and
\beq
\frac{\d\bar Q_{\omega,t}}{\d Q_{\omega,t}}(s)=\e^{-s}.
\label{str-2-quant}
\eeq
\een
\end{theorem}

The relations~\eqref{qes-1-gen} and~\eqref{str-2-quant} are known as the finite
time quantum fluctuation relations.\footnote{They are also sometimes called the
Evans--Searles quantum fluctuation relations.} Theorem~\ref{gen-ttm-quant} was
essentially proven in~\cite[Theorem 7]{Tasaki2003}. For the reader convenience
and future reference, we provide its proof in Section~\ref{sec-proof-1}.

We now return to general $\nu\in\cN$. Our first result is an immediate
consequence of Theorems~\ref{thm-ret-erg} and~\ref{gen-ttm-quant-1-n}:
\begin{theorem}\label{main-erg}
Suppose that the  system $(\cO,\varsigma_\omega,\omega)$ is ergodic.
Then for all $t\in\rr$ and $\nu\in\cN$,
\[
Q_{\nu,t}=Q_{\omega,t}.
\]
\end{theorem}
The above theorem applies to open quantum systems with directly coupled
reservoirs and its assumption holds iff each reservoir system
$(\cO_j,\tau_j,\omega_j)$ is ergodic. In what follows we consider open
quantum systems featuring a small system $\cS$. For any $\nu\in\cS_\cO$
we denote  by $\nu_{\cS}$ the restriction of $\nu$ to $\cO_\cS$.\footnote{For
$A\in\cO_\cS$, $\nu_{\cS}(A)=\nu(A\otimes \one)$.}

\begin{theorem}\label{han-han-n}
Consider an open quantum system where the reservoirs
$\cR_1,\ldots,\cR_M$ are coupled through the small system $\cS$,
each reservoir subsystem $(\cO_j,\tau_j,\omega_j)$ being ergodic.
Let $\nu\in\cN$.
\ben
\item For all $\alpha\in\i\rr$,
\begin{align*}
\fF_{\nu,t}(\alpha)&=\nu_\cS\otimes\omega_\cR\left([D{\omega_{-t}}:D{\omega}]_\alpha\right)\\[1mm]
&=\langle\Omega_{\nu_\cS\otimes\omega_\cR},\Delta_{\omega_{-t}|\omega}^\alpha\Omega_{\nu_\cS\otimes\omega_\cR}\rangle.
\end{align*}
In particular, $Q_{\nu,t}$ is the spectral measure of $-\log\Delta_{\omega_{-t}|\omega}$
for the vector $\Omega_{\nu_\cS\otimes\omega_\cR}$.
\item The measure $Q_{\nu,t}$ is absolutely continuous with respect to $Q_{\omega,t}$ and
\beq
\frac{\d Q_{\nu,t}}{\d Q_{\omega,t}}\leq\dim\,\cK_\cS.
\label{est-1}
\eeq
If $\nu_\cS$ is faithful and $\gamma$ is its smallest eigenvalue, then also
\beq
\gamma\,\dim\,\cK_\cS\leq\frac{\d Q_{\nu,t}}{\d Q_{\omega,t}}.
\label{est-2}
\eeq
\een
\end{theorem}

We equip $\cS_\cO$ with the weak$^\ast$-topology and the set $\cP(\rr)$
of all Borel probability measures on $\rr$ with the weak topology.
By~\cite[Lemma 2.1]{Takenouchi1955} and~\cite[Theorem 1.1]{Fell1960}, the set
of $\omega$-normal states $\cN$ is dense in $\cS_\cO$. This gives
that under the assumptions of either Theorem~\ref{main-erg}
or~\ref{han-han-n}, the map
\beq
\cN\ni\nu\mapsto Q_{\nu,t}\in\cP(\rr)
\label{s-s-sa}
\eeq
uniquely extends to a continuous map
\[
\cS_\cO\ni\nu\mapsto Q_{\nu,t}\in\cP(\rr).
\]
This continuous extension defines $Q_{\nu,t}$ for all $\nu\in\cS_\cO$.
In the case of Theorem~\ref{main-erg}, obviously $Q_{\nu,t}=Q_{\omega,t}$
for all $\nu$. In the case of Theorem~\ref{han-han-n}, $Q_{\nu,t}$ is again
the spectral measure of $-\log\Delta_{\omega_{-t}|\omega}$ for the vector
$\Omega_{\nu_\cS\otimes\omega_\cR}$ and Part~(2) holds. We summarize:
\begin{theorem}\label{thur-sunny}
Consider a non-$\omega$-normal state $\nu\in\cS_\cO$. Then,
Theorems~\ref{main-erg} and~\ref{han-han-n} hold, with $Q_{\nu,t}$ being the
above mentioned continuous extension.
\end{theorem}

\subsection{Remarks}
\label{sec-remarks}

\noindent{\bf 1. Thermodynamic limit.} Thermodynamic limit (abbreviated TDL)
plays a distinguished role in statistical mechanics. It realizes infinitely
extended systems through a limiting procedure involving only finite quantum
systems and is central for the identification of physically relevant objects
in the infinite setting. The precise way the TDL is taken depends on the
structure of the specific physical model under consideration, and often
different approximation routes are possible. This topic is well-understood and
discussed in many places in the literature; see, for example,
\cite{Bratteli1987,Bratteli1981,Ruelle1969} and~\cite[Chapter 5]{Jaksic2010b}.
Since early days, it is known that the modular structure is stable under
TDL~\cite{Araki1974b,Araki1975/76,Araki1976,Araki1977}, and this fact plays
an important role in the foundations of quantum statistical mechanics. The
customary route in discussions of the structural theory is the following:

\bigskip
\noindent {\em Step 1.} A physical notion, introduced in the context of finite
quantum systems, is expressed in a modular form, and through this form is
directly extended, by definition, to a general $C^\ast$/$W^\ast$-dynamical
system. One basic example of such procedure is the introduction of the
KMS-condition as characterization of thermal equilibrium states. This is the
approach we have taken in this work in the introduction of $Q_{\nu,t}$.

\medskip
\noindent {\em Step 2.} In concrete physical models the definitions of
Step~1 are justified by the TDL limit.

\medskip
\noindent Step~2 has been extensively studied in early days of quantum
statistical mechanics and the wealth of obtained results make its implementation
in modern literature most often a routine exercise. For this reason this step is
often skipped. There is rarely a need for making an exception to this rule,
but one is, we believe, in the context of our work. The mixture of quantum
measurements and thermodynamics, which is central to the definition of two-times
quantum measurement entropy production, has a number of unexpected features
that, we believe, require the TDL justification to be put on  solid physical
grounds. More precisely, the formula~\eqref{st-cyr}, which gives the modular
characterization of the initial state decoherence induced by the first
measurement, also allows to postulate this decoherence effect for infinitely
extended systems. The physical and mathematical implications of this Step~1
definition  make its TDL justification necessary.

In the forthcoming article~\cite{Benoist2024b} we will carry out Step~2 for
the 2TMEP of two paradigmatic models in non-equilibrium quantum statistical
mechanics: open quantum spin system on a lattice\footnote{See~\cite{Ruelle2001}.}
and open Spin-Fermion Model\footnote{See the seminal works~\cite{Davies1974,Spohn1978b}.}.

\bigskip
\noindent{\bf 2. The effect  of the first measurement.} The somewhat striking
rigidity of the two-times measurement entropy production statistics that follows
from Theorems~\ref{main-erg} and~\ref{han-han-n}  can be understood in terms of
the dominating effect of the first decoherence inducing measurement. In the TDL,
this effect is dramatic as the measurement induced decoherence forces the
(ergodic) reservoirs into their unique invariant state, while the state of the
small system  $\cS$ has  a marginal effect, being finite dimensional.
In mathematical terms, the first measurement decoherence corresponds to a
projection on $\Ker\log\Delta_\omega$ (see the formula~\eqref{nuplus} below),
which, for infinitely extended systems and due to the ergodicity assumptions of
Theorems~\ref{main-erg} and~\ref{han-han-n}, is finite dimensional.\footnote{In
the case of Theorem~\ref{main-erg} this kernel has dimension~$1$. In the case of
Theorem~\ref{han-han-n} its dimension is $(\dim\,\cK_\cS)^2$.} Along the
sequence of TDL approximations the size of this kernel grows to infinity
together with the dimension of the reservoir Hilbert spaces   to suddenly
drastically shrink  in the limiting state. This  dimension reduction is at the
core of the need for the TDL justification of the Step~1 definition of the
2TMEP.

A proposal for a less invasive measurement protocol through an auxiliary
two-dimensional quantum system (ancilla), that we call Entropic Ancilla State
Tomography, avoids the above  decoherence effect. This  alternative protocol and
its stability are  the topics of~\cite{Benoist2024}, see also Remark~4 below.

\bigskip
\noindent{\bf 3. On the extension to $\nu\in\cS_\cO$.} In
Theorem~\ref{thur-sunny}, the 2TMEP $Q_{\nu,t}$ of $\nu\not\in\cN$ is
defined by the continuous extension of the map~\eqref{s-s-sa} that builds on the
stability results of Theorems~\ref{main-erg} and~\ref{han-han-n}. For
$\nu\not\in\cN$ the representation~\eqref{emm-n} fails (with
$\fF_{\nu,t}(\alpha)$ defined by~\eqref{han-ajde}). In the same vein,
the direct TDL limit justification of $Q_{\nu,t}$ is not possible for
$\nu\not\in\cN$. It is replaced by the TDL limit justification of the
approximating sequence $Q_{\nu_n,t}$, $\nu_n\in\cN$, $\nu_n\to\nu$,
with $\nu_n$'s chosen to reflect the physics of the limiting $\nu$. We
emphasize that when $\nu\notin\cN$, $Q_{\nu,t}$ is defined by first taking
the limit $R\to\infty$ in~\eqref{emm-n} for a fixed normal approximation $\nu_n$
of $\nu$, and then by taking the limit $\nu_n\to\nu$. Taken in reverse order,
this double limit does not necessarily exist, and will not produce the same
limiting measure in general.

\bigskip
\noindent{\bf 4. NESS and quantum Gallavotti--Cohen Fluctuation Theorem.}
This is the topic of the continuation of this work~\cite{Benoist2024}
whose starting points are Theorems~\ref{gen-ttm-quant-1-n}--\ref{thur-sunny},
and we limit ourselves here to a brief comment. The quantum Evans--Searles and
Gallavotti--Cohen Fluctuation Theorems deal with quantum extensions of the
celebrated works~\cite{Evans1994,Gallavotti1995b,Gallavotti1995c} in classical
statistical mechanics, see also the review~\cite{Jaksic2011}. The quantum
Evans--Searles Fluctuation Theorem concerns the Large Deviation Principle (LDP)
for the family of probability measures $(Q_{\omega,t}(t \,\cdot\,))_{t>0}$ in
the limit $t\uparrow\infty$. In parallel with the classical theory, any putative
quantum Gallavotti--Cohen Fluctuation Theorem should concern the entropic
fluctuations with respect to the Non-Equilibrium Steady State (NESS) that the
system $(\cO,\tau,\omega)$ reaches in the large time limit $t\uparrow\infty$.
This NESS is defined as the weak$^\ast$-limit
$\omega_+=\lim_{t\to\infty}\omega_t$.\footnote{The existence of this
limit is typically a deep dynamical problem.} In typical non-equilibrium
setting $\omega_+ \not\in\cN$, and for a long time it was unclear how to
define the statistics $Q_{\omega_+,t}$.
Theorems~\ref{main-erg}--\ref{thur-sunny} provide a route to this definition
that, together with the TDL justification of $Q_{\omega_T,t}$ for all $T,t\geq0$,
is both physically and mathematically natural. However, due to a quantum
decoherence effect, this route comes with a degree of stability that has no
classical counterpart and that identifies the two Fluctuation Theorems under
very general ergodic assumptions that are satisfied in paradigmatic models of
open quantum systems. This triviality aspect is addressed in~\cite{Benoist2024}
by the introduction of Entropic Ancilla State Tomography that we have already
mentioned in Remark~2. Entropic Ancilla State Tomography provides a novel
physical and mathematical perspective on the entropic fluctuations in quantum
statistical mechanics and links the two quantum Fluctuation Theorems in a
non-trivial way.

\bigskip
\noindent{\bf 5. Repeated two-times measurement protocol.} Besides the dynamical
system approach to classical entropy production in which the reference state
plays a central  role (see the  review~\cite{Jaksic2011} for a discussion of
this point), an altogether different random path approach has been developed
in~\cite{Kurchan1998,Lebowitz1999,Maes1999} that does not make use of the
reference state and is applicable to stochastic processes. Its quantum
formulation in the setting of repeated quantum measurement processes goes back
to~\cite{Crooks2008}, and was elaborated in~\cite{Benoist2017c,Benoist2021}.
The advent of experimental methods in cavity and circuit QED, and in particular
the experimental breakthroughs of the Haroche--Raimond and Wineland
groups~\cite{Haroche2013,Haroche2006,Wineland2013} make this complementary
approach particularly relevant. We postpone the comparative discussion of the
two approaches to the forthcoming review.

\bigskip
\noindent {\bf 6. On the choice of $\omega_\cS$.} The proof of
Theorem~\ref{han-han-n} makes explicit use of the special form~\eqref{s-choice}
of $\omega_\cS$, and that has the effect on the values of the constants in the
estimates~\eqref{est-1} and~\eqref{est-2}. However, if $\nu_1$ and $\nu_2$ are
any two states in $\cN_\omega$ such that the restrictions $\nu_{1\cS}$ and
$\nu_{2\cS}$ are faithful with the smallest eigenvalues $\gamma_1$ and
$\gamma_2$, then the  chain rule
\[
\frac{\d Q_{\nu_1,t}}{\d Q_{\nu_2,t}}=\frac{\d Q_{\nu_1,t}}{\d Q_{\omega,t}}
\frac{\d Q_{\omega,t}}{\d Q_{\nu_2,t}}
\]
and~\eqref{est-1}, \eqref{est-2}, give that
\[
\gamma_1\leq \frac{\d Q_{\nu_1, t}}{\d Q_{\nu_2, t}}\leq \frac{1}{\gamma_2}.
\]

\bigskip
\noindent{\bf 7. On the definition of $\fF_{\nu,t}$.} In the context of Theorem~\ref{gen-ttm-quant-1-n},
one also has that
\beq
\fF_{\nu,t}(\alpha)=\lim_{R\to\infty}\frac{1}{R}\int_0^R
\nu\left(\varsigma_\omega^\theta\left([D\omega_{-t}:D\omega]_{\frac{\bar\alpha}{2}}^\ast
[D\omega_{-t}:D\omega]_{\frac{\alpha}{2}}\right)\right)\d\theta.
\label{toul-late}
 \eeq
For finite quantum systems~\eqref{toul-late} is an immediate consequence of the
relation $[\omega,\bar\nu]=0$. In the general case, \eqref{toul-late} is
established by a simple modification of the proof of
Theorem~\ref{gen-ttm-quant-1-n}. In the continuation of this
work~\cite{Benoist2024} we will make use of~\eqref{toul-late} in the context of
the Entropic Ancilla State Tomography.


\section{Proofs}
\label{sec-proofs}

We shall need the following results.

\bel\label{lem:Christmas}
For any $t\in\rr$, one has
$$
\e^{\i t\cL}\Delta_{\omega|\omega_t}\e^{-\i t\cL}=\Delta_{\omega_{-t}|\omega}.
$$
\eel

\noindent{\bf Proof.}
For $A\in\fM$ and $t\in\rr$, we have, taking into account the fact that $J\cL+\cL J=0$,
\begin{align*}
\e^{\i t\cL}\Delta_{\omega|\omega_t}^\frac12\e^{-\i t\cL}A\Omega
&=\e^{\i t\cL}\Delta_{\omega|\omega_t}^\frac12\e^{-\i t\cL}A\e^{\i t\cL}\e^{-\i t\cL}\Omega\\
&=\e^{\i t\cL}J J\Delta_{\omega|\omega_t}^\frac12\tau^{-t}(A)\Omega_t\\
&=\e^{\i t\cL}J\tau^{-t}(A^\ast)\Omega\\
&=\e^{\i t\cL}J\e^{-\i t\cL}A^\ast\e^{\i t\cL}\Omega\\
&=JA^\ast\Omega_{-t}\\
&=\Delta_{\omega_{-t}|\omega}^\frac12 A\Omega,
\end{align*}
where we used that $\e^{\i t\cL}J= J\e^{\i t\cL}$.\hfill\qed

\bep\label{prop:UTRI}
Suppose that $\omega$ is time-reversal invariant with respect to the
time-reversal $\Theta$. Then, there exists an anti-unitary
involution $U$ on $\cH$ such that:
\ben
\item For any $A\in\fM$, $\Theta(A)=UAU^\ast$.
\item $U\Omega=\Omega$ and  $U\cH^+=\cH^+$.
\item $[U,J]=0$.
\item $[U,\cL]=0$.
\item $U^\ast\Delta_{\omega_{-t}|\omega}U=\Delta_{\omega_t|\omega}$ for any $t\in\rr$.
\een
\eep
\noindent{\bf Proof.}
The existence of $U$ as well as Parts~(1)--(3) follow from a simple adaptation of the proof of the
corresponding statements of~\cite[Corrolary~2.5.32]{Bratteli1987}.

To prove Part~(4), we start with the relation $\tau^t\circ\Theta=\Theta\circ\tau^{-t}$,
which yields that
$$
\e^{\i t\cL}UAU^\ast\e^{-\i t\cL}=U\e^{-\i t\cL}A\e^{\i t\cL}U^\ast
$$
for any $t\in\rr$ and $A\in\fM$. It follows that
$$
(U^\ast\e^{\i t\cL}U)A(U^\ast\e^{-\i t\cL}U)=\e^{-\i t\cL}A\e^{\i t\cL},
$$
and since Part~(2) gives $U^\ast\e^{\i t\cL}U\cH^+\subset\cH^+$,
the unicity of the standard Liouvillean yields that
$$
\e^{-\i tU^\ast\cL U}=U^\ast\e^{\i t\cL}U=\e^{-\i t\cL},
$$
from which~(4) follows.

By Parts~(1--2), for any $t\in\rr$ and $A\in\fM$, one has
\begin{align*}
U^\ast\Delta_{\omega_{-t}|\omega}^\frac12UA\Omega
&=U^\ast\Delta_{\omega_{-t}|\omega}^\frac12\Theta(A)\Omega\\[4pt]
&=U^\ast J\Theta(A)^\ast\e^{\i t\cL}\Omega\\[4pt]
&=(U^\ast JU)A^\ast (U^\ast\e^{\i t\cL}U)U^\ast\Omega.
\end{align*}
Invoking Parts~(3--4) further gives
$$
U^\ast\Delta_{\omega_{-t}|\omega}^\frac12UA\Omega
=JA^\ast\e^{-\i t\cL}\Omega=JA^\ast \Omega_t
=\Delta_{\omega_t|\omega}^\frac12A\Omega,
$$
which yields Part~(5).\hfill\qed

\subsection{Proof of Theorem~\ref{gen-ttm-quant-1-n}.}

In parts, the arguments follow closely the proof of Theorem~3.3, Part~(3)
in~\cite{Aschbacher2006}. We give details for the reader's convenience.

We consider first a state  of the form $\nu_B(\cdot)=(B\Omega,\,\cdot\;B\Omega)$
where $B\in\pi(\cO)^\prime$ and $\|B\Omega\|=1$. Then, for any $A\in\cO$,
\begin{align*}
\frac{1}{R}\int_0^R\nu_B(\varsigma_\omega^\theta(A))\d\theta
&=\frac{1}{R}\int_0^R\langle B\Omega,\e^{\i\theta\log\Delta_\omega}A\e^{-\i\theta\log\Delta_\omega}B\Omega\rangle
\d\theta\\[2mm]
&=\frac{1}{R}\int_0^R\langle B^\ast B\Omega,\e^{\i\theta\log\Delta_\omega}A\Omega\rangle\d\theta.
\end{align*}
The von~Neumann ergodic theorem gives that, for all $A\in\cO$,
\[
\nu_{B+}(A)=\lim_{R\to\infty}\frac{1}{R}\int_0^R\nu_B(\varsigma_\omega^\theta(A))\d\theta
=\langle B^\ast B\Omega,PA\Omega\rangle,
\]
where $P$ is the orthogonal projection on $\Ker\log\Delta_\omega$. In particular,
\[
\nu_{B+}([D\omega_{-t}: D\omega]_{\alpha})=\langle B^\ast B\Omega, P\Delta_{\omega_{-t}|\omega}^{\alpha}\Omega\rangle.
\]
This proves that for $\alpha\in\i\rr$,
\beq
\fF_{\nu_B,t}(\alpha)=\lim_{R\to\infty}\frac{1}{R}\int_0^R
\nu_B(\varsigma_\omega^{\theta}([D\omega_{-t}:D\omega]_{\alpha}))\d\theta
=\langle B^\ast B\Omega,P\Delta_{\omega_{-t}|\omega}^{\alpha}\Omega\rangle.
\label{nuplus}
\eeq
The function $\i\rr\ni\alpha\mapsto\fF_{\nu_B,t}(\alpha)$ is continuous.
Moreover, this function is also positive-definite since, for
$z_1,\cdots,z_N\in\cc$ and $\alpha_1,\cdots,\alpha_N\in\i\rr$,
\begin{align*}
\sum_{k,l=1}^N &\fF_{\nu_B,t}(\alpha_k-\alpha_l)z_k\bar z_l
=\left\langle B^\ast B\Omega,P
\left[\sum_{k=1}^Mz_k\Delta_{\omega_{-t}|\omega}^{\alpha_k}\right]^\ast
\left[\sum_{l=1}^Nz_l\Delta_{\omega_{-t}|\omega}^{\alpha_l}\right]\Omega\right\rangle\\[1mm]
&=\left\langle B^\ast B\Omega,P
\left[\sum_{k=1}^N z_k\Delta_{\omega_{-t}|\omega}^{\alpha_k}\Delta_\omega^{-\alpha_k}\right]^\ast
\left[\sum_{l=1}^N z_l\Delta_{\omega_{-t}|\omega}^{\alpha_l}\Delta_\omega^{-\alpha_l}\right]\Omega\right\rangle\\[2mm]
&=\lim_{R\to\infty}\frac{1}{R}\int_0^R\left\langle B^\ast B\Omega,\varsigma_\omega^{\theta}\left(
\left[\sum_{k=1}^Nz_k[D\omega_{-t}:D\omega]_{\alpha_k}\right]^\ast
\left[\sum_{l=1}^Nz_l[D\omega_{-t}:D\omega]_{\alpha_l}\right]\right)\Omega\right\rangle\d\theta\\[2mm]
&=\lim_{R\rightarrow \infty}\frac{1}{R}\int_0^R \nu_B\circ\varsigma_\omega^{\theta}\left(
\left[\sum_{k=1}^N z_k[D\omega_{-t}:D\omega]_{\alpha_k}\right]^\ast
\left[\sum_{l=1}^N z_l[D\omega_{-t}:D\omega]_{\alpha_l}\right]
\right)\d\theta
\geq 0.
\end{align*}
Hence, by the Bochner-Khinchine theorem, there exists unique Borel probability measure $Q_{\nu_B,t}$ on $\rr$ such that,
for all $\alpha\in\i\rr$,
\[
\fF_{\nu_B,t}(\alpha)=\int_\rr\e^{-\alpha s}\d Q_{\nu_B,t}(s).
\]
Let now $\nu$ be an arbitrary $\omega$-normal state on $\cO$. Since $\Omega$ is a cyclic
vector for $\pi(\cO)^\prime$, for any $n\in\nn$ there exists $B_n\in\pi(\cO)'$ such that
\[
\|\nu- \nu_{B_n}\|\leq\frac1n.
\]
This gives that the sequence $\nu_{B_n}$ is Cauchy in norm. If $\nu_+$ is any limit point of the net
\beq
\frac{1}{R}\int_0^R\nu\circ  \varsigma_\omega^\theta \d\theta
\label{net-pr}
\eeq
as $R\uparrow\infty$, we have that
$$
\|\nu_+-\nu_{B_n+}\|\leq\|\nu-\nu_{B_n}\|\le\frac1n.
$$
It follows that $\nu_+$ is the norm limit of $\nu_{B_n+}$ and in particular that the
net~\eqref{net-pr} has the unique limit $\nu_+$. This gives that for all $\alpha\in\i\rr$,
\[
\lim_{R\to\infty}\frac{1}{R}\int_0^R\nu(\varsigma_\omega^{\theta}([D\omega_{-t}:D\omega]_{\alpha}))\d \theta
=\lim_{n\to\infty}\nu_{B_n+}([D\omega_{-t}:D\omega]_{\alpha})
=\nu_+([D\omega_{-t}:D\omega]_{\alpha}),
\]
establishing the existence of $\fF_{\nu,t}$. In addition, we have that
for $\alpha\in\i\rr$,
\[
\fF_{\nu,t}(\alpha)=\lim_{n\to\infty}\int_\rr\e^{-\alpha s}\d Q_{\nu_{B_n},t}(s),
\]
and so, by the L\'evy continuity theorem, there exists unique Borel probability
measure $Q_{\nu,t}$ on $\rr$ such that
\[
\fF_{\nu,t}(\alpha)=\int_\rr\e^{-\alpha s}\d Q_{\nu,t}(s).
\]
\hfill\qed

\subsection{Proof of Theorem~\ref{gen-ttm-quant}.}
\label{sec-proof-1}

\noindent{\bf (1)} By definition of the relative entropy we have
$$
\mathrm{Ent}(\omega_t|\omega)=\langle\Omega_t,\log\Delta_{\omega|\omega_t}\Omega_t\rangle.
$$
Lemma~\ref{lem:Christmas} and the functional calculus allow us to write
\[
\mathrm{Ent}(\omega_t|\omega)
=\langle\Omega,\e^{\i t\cL}\log \Delta_{\omega|\omega_t}\e^{-\i t\cL}\Omega\rangle
=\langle \Omega, \log \Delta_{\omega_{-t}|\omega}\Omega\rangle=-\int_\rr s\, \d Q_{\omega, t}(s).
\]

\medskip
\noindent{\bf (2)} Since $\Omega\in\Dom(\Delta_{\omega_{-t}|\omega}^{1/2})$,
\[
\int_\rr\e^{-s}\d Q_{\omega, t}(s)\d s<\infty,
\]
and this implies the stated regularity of $\fF_{\omega,t}$.
Invoking again Lemma~\ref{lem:Christmas}, and using the fact that
$J^\ast\Delta_{\nu|\mu}J=\Delta_{\mu|\nu}^{-1}$, we write
\begin{align*}
\fF_{\omega,t}(\alpha)
&=\langle\Omega,\Delta_{\omega_{-t}|\omega}^{\alpha}\Omega\rangle\\
&=\langle\Omega,\e^{\i t\cL}\Delta_{\omega|\omega_{t}}^{\alpha}\e^{-\i t\cL}\Omega\rangle\\
&=\langle J\Delta_{\omega_{t}|\omega}^{1/2}\Omega, \Delta_{\omega|\omega_{t}}^{\alpha}J\Delta_{\omega_{t}|\omega}^{1/2}\Omega\rangle\\
&=\overline{\langle\Omega,\Delta_{\omega_{t}|\omega}^{1/2}J^\ast\Delta_{\omega|\omega_{t}}^{\alpha}J\Delta_{\omega_{t}|\omega}^{1/2}\Omega\rangle}\\
&=\overline{\langle\Omega,\Delta_{\omega_{t}|\omega}^{1/2}\Delta_{\omega_t|\omega}^{-\bar\alpha}\Delta_{\omega_{t}|\omega}^{1/2}\Omega\rangle}\\
&=\overline{\langle\Omega,\Delta_{\omega_t|\omega}^{1-\bar\alpha}\Omega\rangle},
\end{align*}
which yields the required identity.

\medskip
\noindent{\bf (3)} By Proposition~\ref{prop:UTRI}, the time-reversal map $\Theta$
has a standard anti-unitary implementation $U$ on $\cH$, such that $U\Omega=\Omega$.
It follows from Part~(5) of the above mentioned proposition that
$$
\fF_{\omega,t}(\alpha)=\langle \Omega,\Delta_{\omega_{-t}|\omega}^\alpha\Omega\rangle
=\langle U\Omega,\Delta_{\omega_{-t}|\omega}^\alpha U\Omega\rangle
=\overline{\langle\Omega,U^\ast\Delta_{\omega_{-t}|\omega}^\alpha U\Omega\rangle}
=\overline{\langle \Omega, \Delta_{\omega_t|\omega}^{\bar\alpha} \Omega\rangle}
=\langle \Omega, \Delta_{\omega_t|\omega}^{\alpha} \Omega\rangle.
$$
Thus, time-reversal invariance implies that
$$
\fF_{\omega,t}(\alpha)=\fF_{\omega,-t}(\alpha)
$$
for $0\leq\Re\,\alpha\leq 1$ and $t\in\rr$. Combined with the identity
obtained in Part~(2), this yields the result.

\medskip\noindent{\bf (4)}
It follows from Part~(3) that for $\alpha\in\i\rr$,
\[
\int_{\rr}\e^{-\alpha s}\d Q_{\omega,t}(s)=\int_\rr\e^{-\alpha s}\e^s\d\bar Q_{\omega,t}(s),
\]
which ends the proof.\hfill \qed

\subsection{Proof of Theorem~\ref{han-han-n}.}

Let $(\psi_1,\ldots,\psi_N)$ be an orthonormal basis of $\cK_\cS$ consisting of
eigenvectors of $\nu_\cS$, and set $P_{ij}=|\psi_i\rangle \langle \psi_j|$. For
the GNS-representation $(\cH_\cS,\pi_\cS,\Omega_\cS)$ of $\cO_\cS$ induced by
the state $\omega_\cS$ given by~\eqref{s-choice} we take
$$
\cH_\cS=\cK_\cS\otimes\cK_\cS,\qquad
\pi_\cS(A)=A\otimes\one,\qquad
\Omega_\cS=\frac1{\sqrt N}\sum_{i=1}^N\psi_i\otimes\psi_i.
$$
If $(\cH_\cR,\pi_\cR,\Omega_\cR)$ is the GNS-representation of $\cO_\cR$ induced by $\omega_\cR$, then
\[
\cH=\cH_\cS\otimes\cH_\cR,\qquad
\pi=\pi_\cS\otimes\pi_\cR,\qquad
\Omega=\Omega_\cS\otimes\Omega_\cR,
\]
and
\[
\log\Delta_\omega=\log\Delta_{\omega_\cR}=\sum_{j=1}^M\log\Delta_{\omega_j}.
\]
By our ergodicity assumption, it follows from Theorem~\ref{thm-li}(2)
and the fact that the Liouvillean of the $j$-th reservoir is $-\beta_j^{-1}\log\Delta_{\omega_j}$),
that $\Ker\log\Delta_\omega$ is spanned by the family
$(\psi_i\otimes  \psi_j\otimes \Omega_\cR)_{1\leq i,j\leq N}$.

We follow up on the proof of Theorem~\ref{gen-ttm-quant-1-n}. With $\nu_{B_n}$
and $P$ as in that proof, we have that
\beq
\begin{split}
{\mathfrak F}_{ \nu_{B_n}, t}(\alpha)&=\langle B_n^\ast B_n\Omega, P\Delta_{\omega_{-t}|\omega}^{\alpha}\Omega\rangle\\[2mm]
&=\sum_{i,j=1}^N\langle B_n^\ast B_n\Omega, \psi_i \otimes \psi_j \otimes \Omega_\cR\rangle\langle \psi_i \otimes \psi_j \otimes \Omega_\cR,
\Delta_{\omega_{-t}|\omega}^{\alpha}\Omega\rangle.
\end{split}
\label{vac-1}
\eeq
Note that
\[
\psi_i\otimes\psi_j\otimes\Omega_\cR=[P_{ij}\psi_j]\otimes \psi_j\otimes \Omega_\cR
=\sqrt{N}\pi(P_{ij})\Omega,
\]
and  so~\eqref{vac-1} gives  that
$$
\fF_{\nu_{B_n},t}(\alpha)
=\sqrt{N}\sum_{i,j=1}^N\nu_{B_n}(\pi(P_{ij}))\langle\psi_i\otimes\psi_j\otimes\Omega_\cR,
\Delta_{\omega_{-t}|\omega}^{\alpha}\Omega\rangle.
$$
Since $\lim_{n\to\infty}\nu_{B_n}(\pi(P_{ij}))=\nu_\cS(P_{ij})$,
and since by the choice of the $\psi_i$'s, $\nu_\cS(P_{ij})=\lambda_i\delta_{ij}$,
where $\lambda_i$ denotes the eigenvalue of $\nu_\cS$ for $\psi_i$, one has
$$
\fF_{\nu,t}(\alpha)=\lim_{n\to\infty}\fF_{\nu_{B_n},t}(\alpha)
=\sqrt{N}\sum_{i=1}^N\lambda_i\langle\psi_i\otimes\psi_i \otimes \Omega_\cR,
\Delta_{\omega_{-t}|\omega}^{\alpha}\Delta_\omega^{-\alpha}\Omega\rangle.
$$
Expressing $\Omega$ in terms of the $\psi_i$'s and invoking~\eqref{equ:ConDef} further leads to
\begin{align*}
\fF_{\nu,t}(\alpha)
&=\sum_{i,k=1}^N\lambda_i\langle \psi_i \otimes \psi_i \otimes \Omega_\cR,[D\omega_{-t}:D\omega]_{\alpha}\psi_k\otimes \psi_k \otimes \Omega_\cR\rangle\\[1mm]
&=\sum_{i=1}^N\lambda_i\langle\psi_i\otimes\psi_i\otimes\Omega_\cR,
[D{\omega_{-t}}:D{\omega}]_{\alpha}\psi_i\otimes\psi_i\otimes\Omega_\cR\rangle,
\end{align*}
where the last equality follows from the fact that
$[D{\omega_{-t}}:D{\omega}]_{\alpha}\in\fM$. Since
$$
\Omega_{\nu_\cS}=\sum_i\sqrt{\lambda_i}\psi_i \otimes \psi_i
$$
is the vector representative of $\nu_\cS$ in $\cH_\cS^+$, the last identity
allows us to conclude that
\[
\fF_{\nu,t}(\alpha)=\nu_\cS\otimes\omega_\cR([D{\omega_{-t}}:D{\omega}]_{\alpha}).
\]
Finally, since $\Omega_{\nu_\cS}\otimes\Omega_\cR\in\Ker\log\Delta_\omega$,
invoking~\eqref{equ:ConDef} again yields
\[
\fF_{\nu,t}(\alpha)=\langle\Omega_{\nu_\cS}\otimes\Omega_\cR,
\Delta_{\omega_{-t}|\omega}^{\alpha}\Omega_{\nu_\cS}\otimes\Omega_\cR\rangle,
\]
from which we can conclude that $Q_{\nu,t}$ is the spectral measure of
$-\log \Delta_{\omega_{-t}|\omega}$ for the vector $\Omega_{\nu_\cS}\otimes \Omega_\cR$.

\medskip
To prove Part~(2), let $B$ be a self-adjoint element of $\pi_\cS(\cO_\cS)^\prime\otimes\one$
such that $B\Omega=\Omega_{\nu_\cS}\otimes\Omega_\cR$. $B$ is invertible iff $\nu_\cS$ is
faithful. For $\alpha\in\i\rr$,
\beq
\begin{split}
\int_\rr\e^{-\alpha s}\d Q_{\nu,t}(s)
&=\langle B\Omega,\Delta_{\omega_{-t}|\omega}^\alpha\Omega_{\nu_\cS}\otimes\Omega_\cR\rangle\\[1mm]
&=\langle B\Omega,\Delta_{\omega_{-t}|\omega}^\alpha\Delta_{\omega}^{-\alpha}\Omega_{\nu_\cS}\otimes\Omega_\cR \rangle\\[1mm]
&=\langle\Omega,\Delta_{\omega_{-t}|\omega}^\alpha\Delta_{\omega}^{-\alpha}B\Omega_{\nu_\cS}\otimes\Omega_\cR\rangle \\[1mm]
&=\langle\Omega,\Delta_{\omega_{-t}|\omega}^\alpha B\Omega_{\nu_\cS}\otimes\Omega_\cR \rangle,
\end{split}
\label{vac-3}
\eeq
and similarly,  if $\nu_\cS$ is faithful,
\beq
\int_\rr \e^{-\alpha s}\d Q_{\omega, t}(s)=\langle \Omega_{\nu_{\cS}}\otimes \Omega_\cR, \Delta_{\omega_{-t}|\omega}^\alpha B^{-1}\Omega \rangle.
\label{vac-2}
\eeq
The identities~\eqref{vac-3} and the spectral theorem give that the measure
$Q_{\nu,t} $ is absolutely continuous with respect to $Q_{\omega,t}$,
with $\d Q_{\nu,t}/\d Q_{\omega,t}$ equal to the projection of
$B\Omega_{\nu_\cS}\otimes\Omega_\cR$ onto the cyclic subspace
$L^2(\rr,\d Q_{\omega,t})$ generated by $-\log \Delta_{\omega_{-t}|\omega}$ and the vector $\Omega$.
Similarly, \eqref{vac-2} gives that $Q_{\omega, t} $ is absolutely continuous
with respect to $Q_{\nu,t}$, with $\d Q_{\omega,t}/\d Q_{\nu,t}$ equal
to the projection of $B^{-1}\Omega$ onto the cyclic subspace
$L^2(\rr,\d Q_{\nu,t})$ generated  $-\log\Delta_{\omega_{-t}|\omega}$ and
the vector $\Omega_{\nu_\cS}\otimes\Omega_\cR$. It remains to prove the
estimates~\eqref{est-1} and~\eqref{est-2}.

By the definition of $\Omega_{\nu_\cS}$ we  have that for $\alpha\in\i\rr$,
\beq
\begin{split}
\langle\Omega_{\nu_\cS}\otimes\Omega_\cR,\Delta_{\omega_{-t}|\omega}^\alpha\Omega_{\nu_\cS}\otimes \Omega_\cR\rangle
&=\sum_{i=1}^N\lambda_i\langle \psi_i\otimes\psi_i\otimes \Omega_\cR,
\Delta_{\omega_{-t}|\omega}^\alpha\Delta_{\omega}^{-\alpha}\psi_i\otimes\psi_i\otimes\Omega_\cR\rangle\\[1mm]
&=\sum_{i=1}^N\lambda_i\langle \psi_i\otimes\psi_i\otimes\Omega_\cR,
\Delta_{\omega_{-t}|\omega}^\alpha\psi_i\otimes\psi_i\otimes\Omega_\cR\rangle.
\end{split}
\label{fl-co}
\eeq
Since for $\epsilon >0$ and $x\in \rr$,
\[
\frac{1}{2\i}\int_{\i\rr}\e^{-\alpha(s-x)}\e^{-|\alpha|\epsilon}\d\alpha=\frac{\epsilon}{\epsilon^2+(x-s)^2},
\]
it follows from~\eqref{fl-co} that
\[
\begin{split}
& \langle  \Omega_{\nu_{\cS}}\otimes \Omega_\cR, \left[\epsilon^2 +(x+\log \Delta_{\omega_{-t}|\omega})^2\right]^{-1}\Omega_{\nu_{\cS}}\otimes \Omega_\cR\rangle \\[1mm]
&= \sum_{i}\lambda_i \langle \psi_i \otimes \psi_i\otimes \Omega_\cR,   \left[\epsilon^2 +(x+\log \Delta_{\omega_{-t}|\omega})^2\right]^{-1}\psi_i\otimes \psi_i\otimes \Omega_\cR\rangle,
\end{split}
\]
which gives that for all $x\in \rr$ and $\epsilon >0$,
\beq
\frac{\ds\int_\rr \frac{\d Q_{\nu, t}(s)}{\ds\epsilon^2 + (x-s)^2}}
{\ds\int_\rr \frac{\d Q_{\omega, t}(s)}{\ds\epsilon^2 + (x-s)^2}}\leq N,
\label{vac-est-1}
\eeq
and if $\lambda_i \geq \gamma$ for all $i$, that
\beq
N\gamma \leq \frac{\ds\int_\rr \frac{\ds\d Q_{\nu, t}(s)}{\epsilon^2 + (x-s)^2}}
{\ds\int_\rr \frac{\d Q_{\omega, t}(s)}{\ds\epsilon^2 + (x-s)^2}}.
\label{vac-est-2}
\eeq
Since\footnote{This is a well-known result, see for example~\cite[Theorem 11]{Jaksic2006d} for a pedagogical  exposition.}
\[
\frac{\d Q_{\nu, t}}{\d Q_{\omega, t}}(x)=\lim_{\epsilon \downarrow 0}\frac{\ds\int_\rr \frac{\d Q_{\nu, t}(s)}{\ds\epsilon^2 + (x-s)^2}}
{\ds\int_\rr \frac{\d Q_{\omega, t}(s)}{\ds\epsilon^2 + (x-s)^2}},
\]
the estimates~\eqref{est-1} and~\eqref{est-2}  follow from~\eqref{vac-est-1} and~\eqref{vac-est-2}.\hfill \qed

\bigskip
\noindent{\small{\bf Data availability and conflict of interest statements.} Data sharing is not applicable to this article as no new data were created or analyzed in this study. The authors have no competing interests to declare that are relevant to the content of this article.

\bibliographystyle{capalpha}
\bibliography{Stability}
\end{document}